\documentclass[onecolumn,showpacs,nobibnotes,nofootinbib]{revtex4}
\usepackage{amsmath,amssymb}
\usepackage{graphicx}

\newcommand{\N}{{\cal{N}}}
\newcommand{\x}{\mathbf{x}}
\newcommand{\y}{\mathbf{y}}
\newcommand{\z}{\mathbf{z}}
\newcommand{\vk}{\mathbf{k}}
\newcommand{\vkp}{\mathbf{k}_0}
\newcommand{\vq}{\mathbf{q}}
\newcommand{\vri}{\mathbf{r}}
\newcommand{\vrh}{\boldsymbol{\rho}}
\newcommand{\vrp}{\boldsymbol{\rho}_0}
\newcommand{\cb}{\mathbf{b}}
\newcommand{\hyper}{{}_2{\rm F}_1}
\newcommand{\cc}[1]{\bar{#1}}
\newcommand{\abs}[1]{\left|#1\right|}
\newcommand{\BesselJ}[2]{J_{#1}\left(#2\right)}

\begin{document}

\title{Traveling waves and geometric scaling at non-zero momentum transfer}
\author{C. Marquet}
\email{marquet@spht.saclay.cea.fr}
\author{R. Peschanski}
\email{pesch@spht.saclay.cea.fr}
\author{G. Soyez\footnote{on leave from the fundamental theoretical physics 
group of the University of Li\`ege.}}
\email{gsoyez@spht.saclay.cea.fr}
\affiliation{SPhT \footnote{URA 2306, unit\'e de recherche associ\'ee au CNRS.}, 
CEA Saclay, B\^{a}t. 774, Orme des Merisiers, 
F-91191 Gif-Sur-Yvette Cedex, France}

\begin{abstract}
We extend the search for traveling-wave asymptotic solutions of the non-linear 
Balitsky-Kovchegov (BK) saturation equation to 
non-forward dipole-target amplitudes. Making use of  conformal invariant 
properties of the Balitsky-Fadin-Kuraev-Lipatov  (BFKL) 
kernel, we exhibit traveling-wave solutions in momentum space in the region 
where the momentum transfer $q$ is smaller than the 
characteristic scale $Q$ of the projectile. We prove geometric scaling in the 
variable $Q/q\Omega_s(Y)$ where $\Omega_s(Y)$ has the 
same energy dependence as in the forward analysis.
%The obtained asymptotic wavefront in momentum space is free from problem caused 
% by the perturbative tail seen in impact parameter space. Consequences for 
%phenomenology are drawn.
\end{abstract}

\maketitle

\section{Introduction}\label{sec:intro}

Geometric scaling \cite{Stasto:2000er} is an interesting phenomenological 
feature of high energy deep-inelastic scattering (DIS). It 
is expressed as a scaling property of the virtual photon-proton cross section, 
namely 
$\sigma^{\gamma^*}(Y,Q)=\sigma^{\gamma^*}\!\left({Q}/{Q_s(Y)}\right). $ $Q$ is 
the virtuality of the photon, $Y$ the total rapidity 
and $Q_s(Y)$ an increasing function of $Y$, called the saturation scale 
\cite{endnote17}. On the theory side of the problem, it is 
convenient to work within the QCD dipole picture of DIS \cite{Nikolaev:1991ja}. 
%In the leading logs  approximation of perturbative QCD at high $Y$, the cross 
%section factories as 
%\[ %begin{equation}\label{eq:gammaN}
%\sigma^{\gamma^*}(Y,Q)
% = \int_0^\infty {\rho}\,d\rho\int_0^1 dz\,|\psi(z,{\rho};Q)|^2\, \N(Y,{\rho})\ 
.
%\] %end{equation}
%$\psi(z,{\rho}Q)$ is the photon wave function on a $q\bar q$ dipole of size 
%$\rho$, and $z$ is the longitudinal momentum fraction of the photon carried by 
%the quark. $\N(Y,\rho)$ is the dipole-target forward scattering amplitude.
In this framework, the geometric scaling 
\begin{equation}\label{eq:defscaling}
\N(Y,\rho)=\N\left({\rho}\ {Q_s(Y)}\right)
\end{equation}
appears to be a genuine property of the conveniently-normalised dipole-target 
forward scattering amplitude $\N(Y,\rho)$, where 
$\rho$ is the size of the dipole and $Y$ its rapidity. In fact, it is known 
that, if this amplitude verifies the non-linear 
Balitsky-Kovchegov (BK) saturation equation \cite{Balitsky,Kovchegov}, this 
property is a mathematical consequence of the 
high-energy behaviour of the solution in terms of traveling waves 
\cite{Munier1}. 
More precisely, the equation is shown to admit traveling-wave solutions which 
translate directly in terms of the geometric scaling 
property (\ref{eq:defscaling}). 

However, the BK equation is written for the dipole-target amplitude in the full 
$2$-dimensional transverse coordinate plane. Hence, 
it is supposed to give solutions for the dipole amplitude $\N(Y,\rho,b)$ where 
$b$ is the impact-parameter of the dipole  projectile 
with respect to the target. Saturation as a function of impact parameter has 
been studied either phenomenologically 
\cite{Munier:2001nr}, in the framework of  models \cite {Kowalski:2003hm}, from 
a semi-classical approach \cite{Bondarenko:2003ym}, 
from numerical studies  of the BK equation 
\cite{Golec-Biernat:2003ym,Gotsman:2004ra} or from an analytic point of view 
\cite{Ikeda:2004zp}. However, referring to solutions of the BK equation, the  
perturbative tail in impact parameter shows up fastly 
in the BK evolution \cite{Golec-Biernat:2003ym}. It appears difficult to be 
consistent in a region where the confinement is expected 
to dominate (see discussions in Refs.\cite{Ferreiro:2002kv,Kovner:2001bh}).

In the present paper we want to tackle the problem of non-forward amplitudes in 
the saturation regime from a new point of view, 
motivated by (and by extension of) the mathematical properties of non-linear 
evolution equations, used in Refs.\cite{Munier1} for 
the forward amplitude. In the present work, we shall stick to the framework of 
the BK equation but our method could be extended to 
different cases with the same mathematical features. 

In Refs.\cite{Munier1}, it was shown that the BK equation for $\N(Y,\rho)$ can 
be considered to lie in the same universality class 
than a well-known equation, namely the Fisher or Kolmogorov-Petrovsky-Piscounov 
(F-KPP) equation \cite{KPP}. It has been shown 
\cite{Bramson} that these equations admit asymptotic traveling-wave solutions 
whose physical meaning \cite{Munier1} is nothing else 
than the geometric scaling property.

More generally, if some general features that we shall now point out are 
realized \cite{ebert}, one can infer the existence and the 
form \cite{Munier1} of traveling wave solutions only from the knowledge of the 
linear part of the kernel \cite{brunet}. Let us quote 
these general features, taking as an example the general structure of the F-KPP 
equations. It is known that they are governed by 
three types of terms: a ``diffusion term'', a ``growth term''  and a non-linear 
``damping term''. In particular, suppose the 
``damping term'' to be absent, the equation is restricted to its linear part and 
leads to an exponential rise of the solution with 
``time'' together with diffusion in ``space''. This is indeed characteristic of 
the  BFKL kernel which governs the linear part of 
the BK equation. In more general cases, called ``pulled front'' cases in the 
literature \cite{ebert}, the presence of similar 
ingredients, together with a specific property of initial conditions being  
steep enough to carry along the critical regime of 
traveling waves, induces the existence and the form of traveling wave solutions. 
The key point is that the main features of these 
solutions can be determined only from the knowledge of the solutions of linear 
part of the equation, which is an easier task than 
looking directly for solutions of a non-linear equation.

Our starting point is the knowledge of the exact solutions of the BFKL equation 
for dipole-dipole scattering, which have been obtained using the conformal 
invariance of the BFKL kernel \cite{Lipatov86,NP,Navelet:1997tx,LipatovReview}. 
Our aim is the following: making use of the powerful mathematical properties 
of conformal symmetry, allowing to formulate exact solutions of the BFKL equation 
with impact parameter and/or momentum transfer, we can construct the solutions of 
the linear part of the BK equation. We shall then look for the 
kinematical domain where the ingredients allowing for the existence 
and derivation of traveling waves are present. Namely, 
schematically (these points shall be discussed more completely in the next 
section):
\begin{itemize}
\item A solution of the linear part of the equation with an exponential growth,
\item A non linear damping effect due to unitarity,
\item A steep enough  initial condition.
\end{itemize}
By contrast with  previous approaches of non-forward amplitudes, we shall not 
try to get solutions in the full phase space (which 
after all is not necessarily dominated by saturation effects) but shall 
concentrate on kinematical regions where the mathematical 
requirements for traveling waves  are fulfilled.

The plan of our study is the following. In section \ref{sec:bindep}, we recall 
in detail the derivation of the traveling-wave speed 
({\em i.e.} the saturation-scale energy dependence) and of the wave front ({\em 
i.e.} the amplitude) in the case of the forward 
amplitude. In section \ref{sec:bdep}, we derive the solutions of the non-forward 
BFKL dipole-dipole amplitudes in various 
representations : full coordinate space, a mixed one where the external 
particles are coordinate space dipoles interacting  with a 
given momentum transfer $q,$ and full momentum space where the external 
particles have given momentum $k,k_0.$ In the next section  
\ref{sec:tw}, we construct the solutions of the linear part of the BK equation 
and select the kinematical domains and 
representations in which one meets the abovementionned requirements and we 
derive the corresponding properties. Finally, in section 
\ref{sec:ccl}, we summarise our results and derive their main consequence,  
namely geometric scaling at nonzero transfer. We show 
how it provides a solution to the puzzling problem related to the confinement 
scale. We point out interesting phenomenological 
consequences of our solutions.

\section{Traveling waves in the forward  case}\label{sec:bindep}

\begin{figure}
\begin{center}
\includegraphics{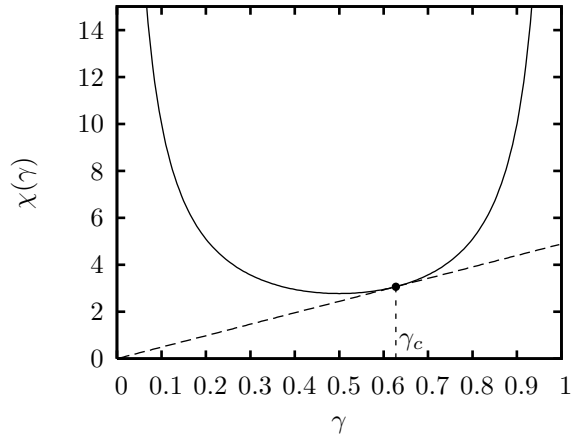}
\end{center}
\caption{Critical exponent $\gamma_c$ for the full BFKL 
kernel.}\label{fig:critqcd}
\end{figure}

Let us recall how traveling waves and geometric scaling emerge from the study of 
the BK equation for forward amplitudes.

In the large-$N_c$ approximation, it is well-known that the high-energy 
behaviour of the dipole forward scattering amplitude off a 
large ({\em e.g.} nuclear) target follows the Balitsky-Kovchegov equation 
\cite{Balitsky,Kovchegov} where we neglect the impact 
parameter dependence. This equation can be put into the form \cite{Kovchegov}
\begin{equation}\label{eq:bkb}
\partial_Y \N(Y,k) = 
\bar{\alpha}\chi(-\partial_L)\N(Y,k)-\bar{\alpha}\N^2(Y,k),
\end{equation}
where 
\[
\chi(\gamma) = 2\psi(1)-\psi(\gamma)-\psi(1-\gamma)
\]
is the BFKL kernel and
\[ %begin{equation}\label{eq:fourier}
\N(Y,k)=\int_0^{\infty} \frac{d\rho}{\rho}J_0(k\rho)\,\N(Y,\rho)
\] %end{equation} 
can be interpreted as the density of gluons in momentum space. Here, $L = 
\log(k^2/k_0^2)$ with $k_0$ some fixed scale. The starting 
point of Ref.\cite{Munier1} is that, if we expand the BFKL kernel to second 
order around $\gamma=\scriptstyle{\frac{1}{2}}$, this 
equation is equivalent, up to a change of variable, to the 
Fisher-Kolmogorov-Petrovsky-Piscounov (F-KPP) equation \cite{KPP}
\begin{equation}\label{eq:KPP}
\partial_t u(t,x) = \partial^2_x u(t,x) + u(t,x) - u^2(t,x)
\end{equation} 
with $t\propto Y$. Looking to the different terms in the right hand side of 
(\ref{eq:KPP}), one can identify  a ``diffusion term'' 
($\partial_x^2 u$), an ``expansion term'' ($u$) and a ``damping term'' ($-u^2$), 
as explained in the introduction.

This equation has been extensively studied and it has been proven that it admits 
traveling-wave solutions \cite{Bramson} {\em i.e.}, 
at large time, the solution can be written 
\[
u(t,x)\underset{t\rightarrow +\infty}{\sim}f(x-m(t))
\]
with 
\[ %begin{equation}\label{eq:velocity}
m(t)=v t- w \log t+ z t^{-1/2} + {\cal O}(1/t)\ ,
\] %end{equation}
where the constants $v,w,z$ can be determined \cite{ebert} from the properties 
of the linear regime. If the initial condition 
behaves like $e^{-\beta x}$ with $\beta > \beta_c = 1$, these coefficients 
acquire critical values, whatever the value of $\beta$ 
is. In particular, the speed $v$ takes the critical value $v_c=2$. The three 
coefficients are the only ``critical'' ones since they 
do not depend on the specific form of the non-linear damping.
% and they were derived \cite{Munier1} for the BK equation.

In Ref.\cite{Munier1}, it has been   shown that this kind of feature is expected 
to be much more general. Let us for instance show  
the properties of the critical regime, when the linear part of the evolution 
admits a superposition of waves for solution. One 
writes:
\begin{equation}\label{eq:waves}
u(t,x) 
 = \int_{c-i\infty}^{c+i\infty} \frac{d\gamma}{2i\pi} u_0(\gamma) e^{-\gamma x + 
\omega(\gamma) t} 
 = \int_{c-i\infty}^{c+i\infty} \frac{d\gamma}{2i\pi} u_0(\gamma) 
e^{-\gamma(x_{wf}+v t) + \omega(\gamma) t},
\end{equation}
where $\omega(\gamma)$ is the Mellin transform of the linear kernel and 
$x_{wf}=x-vt$ is the position relative to the wavefront. 
Then, the non-linear term drive the solution to the critical behaviour 
corresponding to traveling waves at large time
\begin{equation}\label{eq:travwaves}
  u(t,x) \underset{t\to\infty}{\sim} e^{-\gamma_c x_{wf}}.
\end{equation}
The critical exponent $\gamma_c$ corresponds to a wave having a minimal phase 
velocity equal the group velocity (see {\em e.g.} 
\cite{GLR}):
\begin{equation}\label{eq:speed}
v_c = \frac{\omega(\gamma_c)}{\gamma_c} = \left.\partial_\gamma 
\omega(\gamma)\right|_{\gamma_c}.
\end{equation}
The appearance of traveling waves \eqref{eq:travwaves} with critical velocity 
given by \eqref{eq:speed} only depends on a few 
general conditions:
\begin{itemize}
\item $u=0$ is an unstable fix point of the equation, and $u=1$ is a stable fix 
point. 
\item the linearised evolution equation admits solutions of the form 
\eqref{eq:waves}. It generates a growth of the solution and 
non-linearities damp the solution and saturate it. % to 1.
\item The initial condition is steep enough. If $u_0(x) \sim e^{-\gamma_0 x}$, 
this means that we want $\gamma_0 > \gamma_c$.
\end{itemize}

In particular, in the case of the $b$-independent BK equation, $\omega(\gamma) = 
\bar\alpha\chi(\gamma)$ and we find $\gamma_c 
\approx 0.6275$, as represented in Fig.\ref{fig:critqcd}. As noticed in 
\cite{Munier1} we expect $\gamma_0 = 1$ due to QCD colour 
transparency and, therefore, the QCD evolution in rapidity will asymptotically 
reach the traveling wave regime \eqref{eq:travwaves} 
which in turn corresponds to geometric scaling
\begin{equation}\label{eq:scaling}
\N(Y,k) = \N\left(\frac{k^2}{Q_s^2(Y)}\right)
\end{equation}
where the saturation scale $Q_s^2(Y)$ grows as $k_0^2e^{v_cY}$ and the critical 
speed $v_c$ is $4.8834\,\bar\alpha$.

In this paper we investigate in which kinematical regions one has similar 
properties from the BK equation in full phase-space. We 
shall first discuss under which circumstances the BFKL dynamics leads to 
solution of the form \eqref{eq:waves}. We shall then show 
how we can obtain solutions for the linear BK equation and how non linearities 
lead to the formation of traveling waves and to 
geometric scaling.

\section{Linear BFKL dynamics at nonzero momentum transfer}\label{sec:bdep}

If we now take into account the $b$-dependence in the BK equation, we have to 
study the asymptotic solutions of
\begin{equation}\label{eq:bk}
\partial_Y \N(\x,\y) = \frac{\bar{\alpha}}{2\pi}\int d^2z 
\frac{(\x-\y)^2}{(\x-\z)^2(\z-\y)^2} 
\left[\N(\x,\z)+\N(\z,\y)-\N(\x,\y)-\N(\x,\z)\N(\z,\y) \right],
\end{equation}
where $\N(\x, \y)$ is the conveniently-normalised dipole-target scattering 
amplitude. $\x$ and $\y$ are then the transverse space 
coordinates of the quark and antiquark constituting the dipole. This equation 
does not depend explicitly of the target whose centre 
of mass defines the origin. This non-linear equation corresponds to resumming 
QCD fan diagrams in the leading-logarithmic 
approximation \cite{Kovchegov}. 

Our main tool is to use the knowledge of the exact BFKL solutions 
\cite{Lipatov86,NP,Navelet:1997tx} for dipole-dipole scattering. 
We shall study the amplitude $f(\vrh_1, \vrh_2, \vrh_{1'}, \vrh_{2'})$  (see 
Fig.\ref{fig:coord}) where we identify $\vrh_1$ and 
$\vrh_2$ with $\x$ and $\y$ . We impose that one dipole is bigger than the other 
($\rho = \abs{\vrh_1-\vrh_2} \ll \rho_0 = 
\abs{\vrh_{1'}-\vrh_{2'}} \ll 1/\Lambda_{QCD}$).
Indeed, if we want to use the dipole-dipole amplitude in order to study the 
dipole-target amplitude within the BK framework, we need 
to extract solutions of the linear part of the BK equation \eqref{eq:bk} from 
the solutions of the BFKL equation. This is done by 
integrating out the target impact factor, and gives relevant results provided 
the convolution with the large dipole factorises from 
the smaller one. This separation criterium closely depends on the conformal 
invariance properties of the BFKL equation which is one 
of the basic properties of the non-forward linear BFKL dynamics 
\cite{Lipatov86}. 

Since our discussion may lead to different theoretical and phenomenological 
features, we shall perform our analysis in different 
spaces: the coordinate space, the mixed representation where we keep the dipole 
sizes in coordinate space but use the momentum 
transfer $q$ instead of the impact parameter, and the full momentum space.

\subsection{Coordinate space}\label{sec:coord}

\begin{figure}
\includegraphics{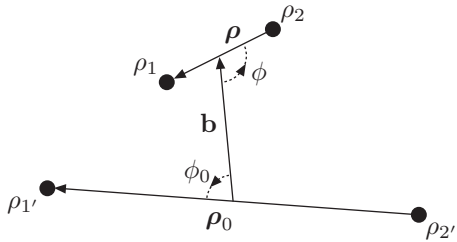}
\caption{Relevant variables in the collision of two dipoles viewed in the 
transverse plane.}\label{fig:coord}
\end{figure}

Let us start from the well-known BFKL amplitude \cite{Lipatov86}
\[ %begin{equation}\label{eq:bfklampl}
{\cal A}(s,q^2) = is \int\frac{d\omega}{2i\pi} e^{\omega Y} f_\omega(q^2),
\] %end{equation}
where $s=e^Y$ is the centre-of-mass energy squared and $q$ is the momentum 
transfer. 
%We keep only the dominant component in energy which corresponds to vanishing 
%conformal spin \cite{Lipatov86}, $n=0$ or $\gamma=\frac{1}{2}+i\nu$),
One writes \cite{Lipatov86}
\begin{eqnarray}
f_\omega(q^2)\delta^{(2)}(\vq-\vq') & = & \int \prod_i 
\frac{d^2\rho_i}{(2\pi)^2}\, 
e^{i\frac{1}{2}(\vrh_1+\vrh_2).\vq}\,\Phi_P(\vrh,\vq) f_\omega(\vrh_1, \vrh_2, 
\vrh_{1'}, 
\vrh_{2'})e^{-i\frac{1}{2}(\vrh_{1'}+\vrh_{2'}).\vq'}\,\Phi_T(\vrp,\vq'), 
\nonumber\\
f_\omega(\vrh_1, \vrh_2, \vrh_{1'}, \vrh_{2'}) & = & \int_{-\infty}^\infty 
\frac{\nu^2\,d\nu}{2\pi(\nu^2+1/4)^2}\frac{1}{\omega-\bar\alpha \chi(\nu)} \int 
d^2r\, 
E^\nu(\vrh_{1}\!-\!\vri,\vrh_{2}\!-\!\vri)\bar E^\nu(\vrh_{1'}\!-\!\vri, 
\vrh_{2'}\!-\!\vri) \label{eq:fomegarrb}
\end{eqnarray}
where 
\[
E^\nu(\vrh_1,\vrh_2)=\left(\frac{|\vrh_1-\vrh_2|}{|\vrh_1||\vrh_2|}\right)^{1+2i
\nu}
\]
are the conformal eigenfunctions of the BFKL kernel with vanishing conformal 
spin quantum number ($n=0$ in the usual terminology 
\cite{Lipatov86}). It corresponds to the dominant contribution at high energy. 
$\Phi_P$ (resp. $\Phi_T$) are the impact factors 
describing the coupling of the BFKL pomeron to the projectile (resp. target), 
depending on the momentum transfer $\vq$ and on the 
dipole sizes $\vrh\! =\! \vrh_1\! -\! \vrh_2$ and 
$\vrp\!=\!\vrh_{1'}\!-\!\vrh_{2'}$.
 
The Dirac distribution which appears in the definition of $f_\omega(q^2)$ comes 
from the fact that, the global system being invariant 
under translation, the expression in the right-hand-side does not depend on 
$\vrh_1\!+\!\vrh_2\!+\!\vrh_{1'}\!+\!\vrh_{2'}$. To be 
more precise, the relevant variables of this problem are the sizes of the two 
dipoles, $\vrh$ and $\vrp$, and the impact parameter $ 
\cb\!=\!\frac{1}{2}(\vrh_1\!+\!\vrh_2\!-\!\vrh_{1'}\!-\!\vrh_{2'})$, as shown on 
Fig.\ref{fig:coord}. Combining these expressions 
and using\footnote{In the following, we shall use indifferently $\gamma$ or 
$\nu$ as superscripts, {\em e.g.} $E^\nu$ and $E^\gamma$ 
denote the same function.} $\gamma=\frac{1}{2}+i\nu$ instead of $\nu$, we obtain 
the following expression for the amplitude
\begin{equation}\label{eq:amplrrb}
{\cal A}(s,q^2) = \frac{is}{(2\pi)^6} \int 
d^2b\,d^2\rho\,d^2\rho_0\,e^{i\vq.\cb}\Phi_P(\vrh, 
\vq)\Phi_T(\vrp,\vq)\int_{\frac{1}{2}-i\infty}^{\frac{1}{2}+i\infty}
\frac{d\gamma}{2i\pi}\,e^{\bar\alpha\chi(\gamma)Y}f^\gamma(\vrh,\vrp,\cb),
\end{equation}
where we have introduced the function $f^\gamma$ as the result of the 
integration over $\vri$ in \eqref{eq:fomegarrb}
\[
f^\gamma(\vrh,\vrp,\cb) = 
\frac{-\left(\gamma-\frac{1}{2}\right)^2}{\gamma^2(1-\gamma)^2}\int 
d^2r\,E^\gamma(\vrh_{1}\!-\!\vri,\vrh_{2}\!-\!\vri)\bar 
E^\gamma(\vrh_{1'}\!-\!\vri, \vrh_{2'}\!-\!\vri).
\]
Using the complex representation for the two-dimensional vectors\footnote{The 
complex representation of the vector $\x=(x_1,x_2)$ is 
\[
x=x_1\!+ ix_2, \qquad \bar x=x_1\!-ix_2.
\]
In this section, we shall explicitly use $\abs{x}$ when the modulus of the 
vector has to be considered.\label{foot:cplx}} (in this 
section, we keep bold characters for vectors and ordinary ones for complex 
numbers), it has been shown \cite{LipatovReview, NP} that 
the result of this integration is
\[
f^\gamma(\rho,\rho_0,b) = \frac{c_\gamma}{\gamma^2(1-\gamma^2)} 
\abs{z}^{2\gamma}\hyper(\gamma,\gamma;2\gamma;z)\hyper(\gamma,\gamma;2\gamma;\cc
{z}) + (\gamma \leftrightarrow 1-\gamma),
\]
where $\hyper$ is the Gauss hypergeometric function \cite{math}, the pre-factor
\begin{equation}\label{eq:cgamma}
c_\gamma = \pi \, 
2^{1-4\gamma}\frac{\Gamma(\gamma)\,\Gamma\left(\frac{3}{2}-\gamma\right)}{\Gamma
(1-\gamma)\Gamma\left(\gamma-\frac{1}{2}\right)}\,,
\end{equation}
and
\begin{equation}\label{eq:ratio}
z = \frac{\rho_{12}\rho_{1'2'}}{\rho_{11'}\rho_{22'}} \equiv 
\frac{4\rho\rho_0}{4b^2-(\rho-\rho_0)^2}
\end{equation}
is the anharmonic ratio. This is a remarkable property due to conformal 
invariance that the BFKL pomeron exchange in coordinate 
space depends on $\rho$, $\rho_0$ and $b$ only through the anharmonic ratio $z$.

Since we are interested in the situation where one small dipole of size $\rho$ 
scatters on a larger one of size $\rho_0$, {\em i.e.} 
$\abs{\rho}\ll\abs{\rho_0}$, we can expand $f^\gamma(\rho, \rho_0, b)$ in series 
of $\rho/\rho_0$. In that limit, the hypergeometric function 
goes to 1 and we obtain
\begin{equation}\label{eq:rrb}
f^\gamma(\rho, \rho_0, b) \approx \frac{c_\gamma}{\gamma^2(1-\gamma)^2} 
\abs{\frac{4\rho\rho_0}{4b^2-\rho_0^2}}^{2\gamma} + (\gamma 
\leftrightarrow 1-\gamma).
\end{equation}
Before going any further, one has to point out that the corrections to this 
expression are of order $(\rho/\rho_0)^2$ but reduce to 
$(\rho/\rho_0)^4$ if we integrate over the phase $\phi$ of $\rho$ (see 
Fig.\ref{fig:coord} for the kinematics). 

The expression \eqref{eq:rrb} still depends on the angle $\phi_0$ between 
$\rho_0$ and $b$. Since this is not relevant for 
phenomenological studies, it is interesting to integrate this result over the 
phases $\phi$ and $\phi_0$ of $\rho$ and $\rho_0$ (one 
may assume that $b$ is real). One obtains
\begin{equation}\label{eq:rrb_mean}
\left\langle f^\gamma(\rho,\rho_0,b)\right\rangle_{\phi,\phi_0}
 = \frac{c_\gamma}{\gamma^2(1-\gamma)^2} 
\left(\frac{4\abs{\rho}\abs{\rho_0}}{\abs{\abs{\rho_0}^2-4\abs{b}^2}}
\right)^{2\gamma}P_{\gamma-1}\left(\frac{\abs{\rho_0}^4
+16\abs{b}^4}{\abs{\abs{\rho_0}^4-16 \abs{b}^4}}\right) 
+ (\gamma \leftrightarrow 1-\gamma),
\end{equation}
where %, here, $\rho$, $\rho_0$ and $b$ represent the modulus of the vectors and 
$P_{\gamma-1}(x)$ is the Legendre function of the first kind \cite{math}. 

Finally, if we additionally take $b$ going to zero in this result, we obtain
\begin{equation}\label{eq:rrbb}
\left\langle f^\gamma(\rho,\rho_0,0)\right\rangle_{\phi,\phi_0}
 = \frac{1}{\gamma^2(1-\gamma)^2}\left(c_\gamma 
\abs{\frac{4\rho}{\rho_0}}^{2\gamma}
 + c_{1-\gamma} \abs{\frac{4\rho}{\rho_0}}^{2-2\gamma}\right)
\end{equation}
which exhibits the same power behaviour as in the forward case.

\subsection{Mixed space}\label{sec:mixed}

It is useful to introduce a mixed space representation \cite{Lipatov86} in terms 
of $\rho$, $\rho_0$ and $q$. One obtains from 
\eqref{eq:amplrrb}
\begin{equation}\label{eq:amplrrq}
{\cal{A}}(s,q^2) = \frac{is}{(2\pi)^4}\int d^2\rho\, d^2\rho_0\, \Phi_P(\vrh, 
\vq)\Phi_T(\vrp,\vq)\int_{-\infty}^\infty 
\frac{d\nu}{2\pi}\,f_{q}^\nu(\vrh, \vrp)\,e^{\bar\alpha\chi(\nu)Y}
\end{equation}
where
\begin{equation}\label{eq:fqgamma}
f_q^\nu(\vrh, \vrp) = \int 
\frac{d^2b}{(2\pi)^2}\,e^{i\vq.\cb}\,f^\nu(\vrh,\vrp,\cb) = 
\frac{\abs{\vrh}\abs{\vrp}}{16(\nu^2+1/4)^2} 
\bar E_q^\nu(\vrp) E_q^\nu(\vrh)
\end{equation}
acquires a factorised form with 
\[
E_q^\nu(\vrh)=\frac{4c_\gamma}{\pi^2|\vrh|}\int d^2z\
e^{i\vq.\z}E^\nu\left(\z+\frac{\vrh}{2},\z-\frac{\vrh}{2}\right).
\]
Using again the complex representation for the two-dimensional vectors, one 
finds (see \cite{NP})
\[
     E_q^\nu(\rho)  = \abs{q}^{2i\nu} 2^{-6i\nu} \Gamma^2(1-i\nu)
             \left\lbrack 
	     \BesselJ{-i\nu}{\frac{\bar 
q\rho}{4}}\BesselJ{-i\nu}{\frac{q\bar\rho}{4}} 
	   - \BesselJ{ i\nu}{\frac{\bar q\rho}{4}}\BesselJ{ 
i\nu}{\frac{q\bar\rho}{4}}
	     \right\rbrack \quad\text{and}\qquad
\bar{E}_q^\nu(\rho_0)  = E_q^{-\nu}(\rho_0)\,,
\]
where $\BesselJ{\alpha}{z}$ is the Bessel function of the first kind.
Using $\gamma = \frac{1}{2}+i\nu$ instead of $\nu$, the $(\nu \leftrightarrow 
-\nu)$ symmetry turns into a $(\gamma \leftrightarrow 
1-\gamma)$ symmetry and we finally get
\begin{eqnarray}
f_q^\gamma(\rho, \rho_0) =
  \frac{\abs{\rho\rho_0}}{16 \gamma^2(1-\gamma)^2} 
  \Gamma^2\left(\frac{3}{2}-\gamma\right) 
\Gamma^2\left(\frac{1}{2}+\gamma\right) \label{eq:bfkl_rrq}
&&\left\lbrack 
     \BesselJ{\frac{1}{2}-\gamma}{\frac{\bar q\rho}{4}}
     \BesselJ{\frac{1}{2}-\gamma}{\frac{q\bar \rho}{4}} 
   - \BesselJ{\gamma-\frac{1}{2}}{\frac{\bar q\rho}{4}}
     \BesselJ{\gamma-\frac{1}{2}}{\frac{q\bar \rho}{4}} 
  \right\rbrack\\\times&& \left\lbrack 
     \BesselJ{\gamma-\frac{1}{2}}{\frac{\bar q\rho_0}{4}}
     \BesselJ{\gamma-\frac{1}{2}}{\frac{q\bar \rho_0}{4}} 
   - \BesselJ{\frac{1}{2}-\gamma}{\frac{\bar q\rho_0}{4}}
     \BesselJ{\frac{1}{2}-\gamma}{\frac{q\bar \rho_0}{4}} 
  \right\rbrack\,.\nonumber
\end{eqnarray}
Let us notice that this expression has the crucial feature that it is {\em 
factorised} in $\rho$ and $\rho_0$ by contrast with the 
$b$ representation. This property, inherent to the fact that we use the momentum 
transfer $q$ instead of the impact parameter $b$, 
will be of prime importance for obtaining the solutions of the BK equation.

As explained before, we are interested in the situation where one dipole is much 
larger than the other one ($\rho_0\gg\rho$). In 
order to expand the Bessel functions in series of their respective arguments, we 
shall consider three situations: $\rho_0\gg\rho\gg 
1/q$, $\rho_0\gg1/q\gg\rho$ and $1/q\gg\rho_0\gg\rho$. The results then follow 
from the asymptotic expansion of the Bessel functions
\begin{eqnarray}
\BesselJ{\mu}{z}\BesselJ{\mu}{\bar z}-\BesselJ{-\mu}{z}\BesselJ{-\mu}{\bar z}
& \stackrel{\abs{z}\ll 1}{\longrightarrow} & 
\frac{1}{\Gamma^2(1+\mu)}\abs{\frac{z}{2}}^{2\mu} 
                             - 
\frac{1}{\Gamma^2(1-\mu)}\abs{\frac{z}{2}}^{-2\mu}, \label{eq:BesselSmall}\\
& \stackrel{\abs{z}\gg 1}{\longrightarrow} & \frac{-2}{\pi\abs{z}} \sin(\mu 
\pi)\cos\left[2\Re e(z)\right].\label{eq:BesselLarge}
\end{eqnarray}

Using these expressions we obtain the following results for the three relevant 
cases.
\begin{itemize}
\item \underline{Case 1}: $1/q\gg\rho_0\gg\rho$.\\
Using \eqref{eq:bfkl_rrq} and \eqref{eq:BesselSmall}, we find
\begin{eqnarray*}
f_q^\gamma(\rho,\rho_0) & = & \frac{\abs{\rho\rho_0}}{16 \gamma^2(1-\gamma)^2} 
  \Gamma^2\left(\frac{3}{2}-\gamma\right) 
\Gamma^2\left(\frac{1}{2}+\gamma\right) \label{eq:bfkl_bbq}\\
&\times&\left\lbrack 
  \frac{1}{\Gamma^2\left(\frac{3}{2}-\gamma\right)}\abs{\frac{\rho 
q}{8}}^{1-2\gamma}
- \frac{1}{\Gamma^2\left(\frac{1}{2}+\gamma\right)}\abs{\frac{\rho 
q}{8}}^{2\gamma-1}
\right\rbrack \left\lbrack 
  \frac{1}{\Gamma^2\left(\frac{1}{2}+\gamma\right)}\abs{\frac{\rho_0 
q}{8}}^{2\gamma-1}
- \frac{1}{\Gamma^2\left(\frac{3}{2}-\gamma\right)}\abs{\frac{\rho_0 
q}{8}}^{1-2\gamma}
\right\rbrack.
\end{eqnarray*}
In particular, if we take $q\to 0$ in this expression, we recover
\begin{equation}
f_q^\gamma(\rho,\rho_0) \stackrel{q\to 0}{\longrightarrow} 
\frac{\abs{\rho\rho_0}}{16 
\gamma^2(1-\gamma)^2}\left(\abs{\frac{\rho}{\rho_0}}^{1-2\gamma} + 
\abs{\frac{\rho}{\rho_0}}^{2\gamma-1}\right).
\end{equation}

\item \underline{Case 2}: $\rho_0\gg1/q\gg\rho$.\\
Due to the asymptotic expansion \eqref{eq:BesselLarge}, the $\rho_0$ part will 
not depend on $\gamma$ anymore and we obtain
\begin{equation}\label{eq:rrq_case2}
f_q^\gamma(\rho,\rho_0) = 
\frac{-1}{2\pi}\frac{\cos(\gamma\pi)}{\gamma^2(1-\gamma)^2} \cos\left[\frac{\Re 
e(\rho_0\bar 
q)}{2}\right]\frac{1}{\abs{q}^2} 
\left[\Gamma^2\left(\frac{3}{2}-\gamma\right)\abs{\frac{\rho q}{8}}^{2\gamma}- 
\Gamma^2\left(\frac{1}{2}+\gamma\right)\abs{\frac{\rho 
q}{8}}^{2-2\gamma}\right].
\end{equation}
Note that this amplitude still depends on the angle $\psi_0$ between $q$ and 
$\rho_0$. If we integrate out this angle together with 
the dependence on the angle $\psi$ between $q$ and $\rho$, we have
\begin{equation}\label{eq:rrq_case2_noangle}
\left\langle f_q^\gamma(\rho,\rho_0) \right\rangle_{\psi,\psi_0} = 
\frac{-\cos(\gamma\pi)}{\gamma^2(1-\gamma)^2} 
\cos\left(\frac{\abs{\rho_0 
q}}{2}-\frac{\pi}{4}\right)\sqrt{\frac{1}{\pi^3\abs{\rho_0 q^3}}} 
\left[\Gamma^2\left(\frac{3}{2}-\gamma\right)\abs{\frac{\rho q}{8}}^{2\gamma}- 
\Gamma^2\left(\frac{1}{2}+\gamma\right)\abs{\frac{\rho 
q}{8}}^{2-2\gamma}\right].
\end{equation}

\item \underline{Case 3}: $\rho_0\gg\rho\gg 1/q$.\\
In this case, both the $\rho$- and $\rho_0$-dependent parts of 
\eqref{eq:bfkl_rrq} involve expansion \eqref{eq:BesselLarge} which 
gives
\begin{equation}\label{eq:rrq_case3}
f_q^\gamma(\rho,\rho_0) = 
\frac{-4}{\pi^2}\frac{\cos^2(\gamma\pi)}{\gamma^2(1-\gamma)^2}
\Gamma^2\left(\frac{1}{2}+\gamma\right)\Gamma^2\left(\frac{3}{2}-
\gamma\right)\frac{1}{\abs{q}^2}\cos\left[\frac{\Re e(\rho\bar 
q)}{2}\right]\cos\left[\frac{\Re e(\rho_0\bar q)}{2}\right]
\end{equation}
and, averaging over the angles $\psi$ and $\psi_0$,
\begin{equation}\label{eq:rrq_case3_noangle}
\left\langle f_q^\gamma(\rho,\rho_0) \right\rangle_{\psi,\psi_0} = 
\frac{-16}{\pi^4}\frac{\cos^2(\gamma\pi)}{\gamma^2(1-\gamma)^2}
\Gamma^2\left(\frac{1}{2}+\gamma\right)\Gamma^2\left(\frac{3}{2}
-\gamma\right)\frac{1}{\abs{\rho\rho_0q^4}}
\cos\left(\frac{\abs{\rho q}}{2}-\frac{\pi}{4}\right)
\cos\left(\frac{\abs{\rho_0 q}}{2}-\frac{\pi}{4}\right).
\end{equation}
\end{itemize}

%Before going to the momentum space, we have to note that 
%\eqref{eq:rrq_case2_noangle} and \eqref{eq:rrq_case3_noangle} are not 
%positive defined.

\subsection{Momentum space}\label{sec:mom}

The situation in momentum space is very similar to the one in mixed space. 
If we still neglect the contribution of higher conformal 
spins, we find 
\begin{equation}\label{eq:amplkkq}
{\cal A}(s,q^2) = is \int d^2k\, d^2k_0\, \Phi_P(\vk, \vq)\Phi_T(\vkp, 
\vq)\int_{-\infty}^\infty \frac{d\nu}{2\pi}\, e^{\bar 
\alpha\chi(\nu)Y} f^\nu(\vk, \vkp, \vq),
\end{equation}
where $\Phi_P$ and $\Phi_T$ are the impact factors in momentum space 
\[ %begin{equation}\label{eq:impactkkq}
\Phi_P(\vk, \vq)=\int \frac{d^2\rho}{(2\pi)^2}\, 
e^{-i(\vk-\vq/2).\vrh}\,\Phi_P(\vrh, \vq),
\] %end{equation}
and $f^\nu(\vk,\vkp, \vq)$ is the Fourier transform of $f_q^\nu(\vrh, \vrp)$:
\begin{equation}\label{eq:fkkqdef}
f^\nu(\vk,\vkp,\vq) = \frac{1}{(2\pi)^4}\frac{1}{(\nu^2+1/4)^2}
   \int d^2\rho\, e^{i(\vk-\vq/2).\vrh}\, \frac{\abs{\vrh}}{4} E_q^\nu(\vrh)
   \int d^2\rho_0\, e^{i(\vkp-\vq/2).\vrp}\, \frac{\abs{\vrp}}{4} \bar 
E_q^\nu(\vrp).
\end{equation}
The detailed calculation of this integral is presented in Appendix 
\ref{ap:fourier} and the relation with conformal invariance is 
developed in Appendix \ref{ap:sl2c}. The final expression, using again the 
complex representation (see footnote \ref{foot:cplx}) and 
replacing $\nu$ by $\gamma=\frac{1}{2}+i\nu$, is
\begin{eqnarray}
f^\gamma(k,k_0,q) 
  & = & \frac{-1}{(2\pi)^4}\frac{\sin^2(\gamma\pi)}{16\gamma^2(1-\gamma)^2} 
\Gamma^2\left(\frac{1}{2}+\gamma\right)\Gamma^2\left(\frac{3}{2}-\gamma\right)
        \left(\frac{4}{\abs{kk_0}}\right)^3\nonumber\\
&\times&\left\lbrack 
\frac{\Gamma^2(1+\gamma)}{\Gamma^2\left(\frac{1}{2}+\gamma\right)}
        \abs{\frac{q}{4k}}^{2\gamma-1}
        \hyper\left(\gamma,1+\gamma;2\gamma;\frac{q}{k}\right)
        \hyper\left(\gamma,1+\gamma;2\gamma;\frac{\bar q}{\bar k}\right)
        - (\gamma\leftrightarrow 1-\gamma)
	\right\rbrack\nonumber\\
&\times&\left\lbrack 
\frac{\Gamma^2(1+\gamma)}{\Gamma^2\left(\frac{1}{2}+\gamma\right)}
        \abs{\frac{q}{4k_0}}^{2\gamma-1}
        \hyper\left(\gamma,1+\gamma;2\gamma;\frac{q}{k_0}\right)
        \hyper\left(\gamma,1+\gamma;2\gamma;\frac{\bar q}{\bar k_0}\right)
        - (\gamma\leftrightarrow 1-\gamma)
	\right\rbrack.\label{eq:fkkpq}
\end{eqnarray}
As in the mixed space representation, this expression has the remarkable 
property of being {\em factorised} in $k$ and $k_0$. This 
appears from the fact that we use the momentum transfer $q$ and is lost if we go 
back to impact-parameter $b$.

We can now consider these expressions in the case where one dipole is much harder 
than the other, corresponding to $k\gg k_0$. Thanks 
to the fact that $f^\gamma$ is factorised in $q/k$ and $q/k_0$, we again need to 
consider three situations: $k\gg k_0\gg q$, $k\gg 
q\gg k_0$ and $q\gg k\gg k_0$. When $q$ is small compared to $k$, it is 
sufficient to consider
\[
\hyper\left(\gamma,1+\gamma;2\gamma;\frac{q}{k}\right) \stackrel{\abs{k}\gg 
\abs{q}}{\approx} 1\,.
\]
The behaviour at large $q$ is more complicated. We show in Appendix 
\ref{ap:asymp} that, averaging over the angle $\theta$ between 
$k$ and $q$, we have
\begin{eqnarray}
\left\langle \frac{\Gamma^2(1+\gamma)}{\Gamma^2\left(\frac{1}{2}+\gamma\right)}
\abs{\frac{q}{4k}}^{2\gamma-1}
\hyper\left(\gamma,1+\gamma;2\gamma;\frac{q}{k}\right)
\hyper\left(\gamma,1+\gamma;2\gamma;\frac{\bar q}{\bar k}\right)
- (\gamma\leftrightarrow 1-\gamma)
\right\rangle_{\!\theta}\nonumber\\
\stackrel{\abs{k}\ll \abs{q}}{\longrightarrow} 
4\gamma^2(1-\gamma)^2\frac{\cos(\gamma\pi)}{\sin(\gamma\pi)}\abs{\frac{q}{k}}^{-
3}\,\log\abs{\frac{q}{k}}.\label{eq:2f1exp}
\end{eqnarray}
Before considering the three cases in more details, one may note that the 
$q^{-3}$ behaviour could have been anticipated from the 
fact that, in that limit, the Fourier transform of $E_q^\nu(\rho)$ does not 
depend on $k$. Note however the logarithmic term arising 
from the hypergeometric function.

Let us now consider the results for the three different cases.
\begin{itemize}
\item \underline{Case 1}: $k\gg k_0\gg q$.\\
In this case, both parts of the expression show a power behaviour,
\begin{eqnarray}
f^\gamma(k,k_0,q) & = & \frac{-1}{\pi^4} 
\frac{\sin^2(\gamma\pi)}{4\gamma^2(1-\gamma)^2}\Gamma^2\left(\frac{3}{2}
-\gamma\right)\Gamma^2\left(\frac{1}{2}+\gamma\right) 
\frac{1}{\abs{kk_0}^3}\label{eq:kkq_case1}\\
&\times& \left\lbrack 
   \frac{\Gamma^2\left(1+\gamma\right)}{\Gamma^2\left(\frac{1}{2}+\gamma\right)}
   \abs{\frac{q}{4k}}^{2\gamma-1} 
 - \frac{\Gamma^2\left(2-\gamma\right)}{\Gamma^2\left(\frac{3}{2}-\gamma\right)}
   \abs{\frac{q}{4k}}^{1-2\gamma}\right\rbrack
 \left\lbrack 
   \frac{\Gamma^2\left(1+\gamma\right)}{\Gamma^2\left(\frac{1}{2}+\gamma\right)}
   \abs{\frac{q}{4k_0}}^{2\gamma-1} 
 - \frac{\Gamma^2\left(2-\gamma\right)}{\Gamma^2\left(\frac{3}{2}-\gamma\right)}
   \abs{\frac{q}{4k_0}}^{1-2\gamma}\right\rbrack\nonumber\\
 & = & 
\frac{1}{(2\pi)^2}\frac{1}{\abs{kk_0}^3}
\left\lbrack\abs{\frac{k}{k_0}}^{1-2\gamma}-\frac{\sin^2(\gamma\pi)}
{16\pi^2\gamma^2(1-\gamma)^2}\frac{\Gamma^4(1+\gamma)\Gamma^2\left(\frac{3}{2}
-\gamma\right)}{\Gamma^2\left(\frac{1}{2}+\gamma\right)}\abs{\frac{q^2}
{16kk_0}}^{2\gamma-1}+(\gamma\to 1-\gamma)\right\rbrack.\nonumber
\end{eqnarray}

In the distributed product, we clearly see that, taking the limit $q\to 0$, we 
recover an expression depending only on the ratio 
between the momenta of the two dipoles
\begin{equation}\label{eq:kkq_case10}
f^\gamma(k,k_0,q)\stackrel{q\to 
0}{\longrightarrow}\frac{1}{(2\pi)^2}\frac{1}{\abs{kk_0}^3}\left(\abs{\frac{k}
{k_0}}^{1-2\gamma}+\abs{\frac{k}{k_0}}^{2\gamma-1}
\right).
\end{equation}

\item \underline{Case 2}: $k\gg q\gg k_0$.\\
At intermediate values of $q$, the dependence on $k_0$ goes away and we recover 
an analytic power behaviour in $\gamma$ which, as we 
shall see in the next section, will lead to traveling waves and geometric 
scaling
\begin{eqnarray}
f^\gamma(k,k_0,q) & = & 
\frac{-1}{2\pi^4}\sin(2\gamma\pi)\Gamma^2\left(\frac{3}{2}-\gamma\right)\Gamma^2
\left(\frac{1}{2}+\gamma\right)\nonumber\\
&& \log\abs{\frac{q}{k_0}}\frac{1}{\abs{kq}^3}\left\lbrack 
\frac{\Gamma^2\left(1+\gamma\right)}{\Gamma^2\left(\frac{1}{2}+\gamma\right)}
\abs{\frac{q}{4k}}^{2\gamma-1} - 
\frac{\Gamma^2\left(2-\gamma\right)}{\Gamma^2\left(\frac{3}{2}-\gamma\right)}
\abs{\frac{q}{4k}}^{1-2\gamma} 
\right\rbrack.\label{eq:kkq_case2}
\end{eqnarray}

\item \underline{Case 3}: $q\gg k\gg k_0$.\\
This limit gives
\begin{equation}\label{eq:kkq_case3}
f^\gamma(k,k_0,q) = 
\frac{-4}{\pi^4}\cos^2(\gamma\pi)\Gamma^2\left(\frac{3}{2}-\gamma\right)\Gamma^2
\left(\frac{1}{2}+\gamma\right)\gamma^2(1-\gamma)^2 
\frac{1}{\abs{q}^6}\log\abs{\frac{q}{k_0}}\log\abs{\frac{q}{k}}.
\end{equation}
%Note that this expression is positive since, for $\gamma = \frac{1}{2}+i\nu$, 
%we have $-\cos^2(\gamma\pi) = \sinh^2(\nu\pi)$.

\end{itemize}

\section{Traveling waves at nonzero transfer}\label{sec:tw}

In this section, we investigate the consequences of our formul\ae\ for linear 
BFKL dynamics on asymptotic solutions when the 
non-linear effects of the BK equation are switched on. The key point is that 
linear BFKL dynamics can provide solutions for the 
linear part of the BK equation. Indeed, when factorisation between the target 
and the projectile is fulfilled, the impact factor of 
the target can be integrated out keeping a linear evolution governed by the same 
kernel. It is thus a solution of the linear part of 
the BK equation which should depend only on the kinematic variables of the 
projectile.

In the next step, the asymptotic solutions of the BK equation can be deduced 
from the knowledge of its linear part, using quite 
general arguments which were recalled in Sections \ref{sec:intro} and 
\ref{sec:bindep}. 
For each of the kinematical configurations introduced in the previous section, 
we check whether the conditions for the existence of 
traveling waves are fulfilled and, in these cases, derive their expression.

Let us first concentrate on the full momentum representation \eqref{eq:amplkkq}. 
We define the function
\begin{equation}\label{eq:bkfkq}
f(\vk,\vq) = \int d^2k_0\,\Phi_T(\vk_0,\vq)\int\frac{d\gamma}{2i\pi}\,e^{\bar 
\alpha\chi(\gamma)Y} f^\gamma(\vk,\vk_0,\vq),
\end{equation}
depending only on $\vk$ and $\vq$. It is remarkable that in this representation, 
the exact factorisation property of 
$f^\gamma(\vk,\vk_0,\vq)$ (see \eqref{eq:fkkqdef}) gives rise to the simple 
expression 
\begin{equation}\label{eq:kkqfkq}
f(\vk,\vq) = \int\frac{d\gamma}{2i\pi}\,e^{\bar 
\alpha\chi(\gamma)Y}\phi^\gamma(\vq)\,f^\gamma(\vk,\vq),
\end{equation}
where
\[
\phi^\gamma(\vq) = \int d^2k_0\, \frac{d^2\rho_0}{(2\pi)^2}\, 
e^{i(\vk_0-\vq/2).\vrp}\,\frac{\abs{\vrp}}{4}\bar 
E_q^\gamma(\vrp)\Phi_T(\vk_0,\vq)
\] 
factorises out the target dependence, defining the initial condition, and
\begin{eqnarray}
\lefteqn{f^\gamma(\vk, \vq) 
 = \int \frac{d^2\rho}{(2\pi)^2}\, 
e^{i(\vk-\vq/2).\vrh}\,\frac{\abs{\vrh}}{4}E_q^\gamma(\vrh)}\\
&&= 
\frac{2^{4-6\gamma}\,\Gamma^2\!\left(\frac{3}{2}-\gamma\right)}{(2\pi)^2}
\sin(\gamma\pi)\frac{\abs{q}^{2\gamma-1}}{\abs{k}^3} %\nonumber\\
%& \times & 
\left\lbrack 
\frac{\Gamma^2\!\left(1+\gamma\right)}{\Gamma^2\!\left(\frac{1}{2}+\gamma\right)
}\abs{\frac{q}{4k}}^{2\gamma-1}\hyper\!\left(\gamma+
1,\gamma;2\gamma;\frac{q}{k}\right)\hyper\!\left(\gamma+1,\gamma;2\gamma;
\frac{\bar q}{\bar k}\right)-(\gamma\to 1-\gamma) 
\right\rbrack\nonumber 
\end{eqnarray}
is calculated in Appendix \ref{ap:sl2c} and enters in formula \eqref{eq:fkkpq}.
It is obvious from the structure of \eqref{eq:bkfkq}, a linear superposition of 
eigenfunctions of the BFKL kernel, that it provides 
a natural basis for the solutions of the linear part of the BK equation in 
momentum space.

Therefore, we look for the kinematical domains where $f^\gamma(\vk, \vq)$ takes 
the appropriate exponential behaviour, corresponding 
to the wave superposition property \eqref{eq:waves} discussed in section 
\ref{sec:bindep}. It will lead to traveling-wave solutions 
for the full BK equation.

We observe that, both in the cases 1 and 2 (formul\ae\ \eqref{eq:kkq_case10} and 
\eqref{eq:kkq_case2}), {\em i.e.} in the limit 
$k\!\gg\!q$, the solution of the linear part of the BK equation can be recast, 
using the symmetry $\gamma \leftrightarrow 1-\gamma$, 
under the form
\[
f(\vk,\vq) = \frac{1}{k^2} 
\int\frac{d\gamma}{2i\pi}\,\phi^\gamma(\vq)\,e^{\bar\alpha\chi(\gamma)Y-\gamma 
L},
\]
where
\[
L=\log\left(\frac{k^2}{q^2}\right)
\]
expresses the leading power-like behaviour at large $k/q$ and all remaining 
factors in $\gamma$ and $\vq$ has been reabsorbed in 
$\phi^\gamma(\vq)$. This clearly emphasises the fact that the region of interest 
is $k\!\gg\!q$, independently of $k_0$. By 
contrast, in the case 3 treating the large-$q$ expression, we see that equation 
\eqref{eq:kkq_case3}, though being factorised, does 
not fulfil the same property. We thus do not expect traveling-wave solutions in 
this limit.

Let us now include the effect of the nonlinear term in the BK equation. Due to 
colour transparency ($f(\vk, \vq)\sim k^{-2}$ for 
$k\gg q$), one can apply the general arguments exposed in Section 
\ref{sec:bindep}. Therefore, the solution of the BK equation 
reaches asymptotically the traveling-wave structure and exhibits geometric 
scaling. The saturation scale 
\begin{eqnarray}\label{eq:newqs}
Q_s^2(Y) & = & q^2 \Omega_s^2(Y) \nonumber  \\[-3mm]
&&\\ 
&\sim& q^2  
\exp\left[\bar\alpha\frac{\chi(\gamma_c)}{\gamma_c}Y-\frac{3}{2\gamma_c}\log(Y) 
\right]\nonumber 
\end{eqnarray}
has the same rapidity dependence as in the forward case but is proportional to 
the momentum transfer. The critical exponent 
$\gamma_c$ is still given by equation \eqref{eq:speed}. The asymptotic form of 
the amplitude, {\em i.e.} the wavefront, can be 
written at small $q/k$
\begin{equation}\label{eq:newfront}
f(\vk,\vq) \stackrel{Y\to\infty}{\sim} 
\frac{1}{k^2}\,\phi^{\gamma_c}(\vq)\,\log\left(\frac{k^2}{q^2\Omega_s^2(Y)}
\right)\,\abs{\frac{k^2}{q^2\Omega_s^2(Y)}}^{-\gamma_c}\,
\exp\left[-\frac{1}{2\bar\alpha\chi''(\gamma_c)Y}\log^2\left(
\frac{k^2}{q^2\Omega_s^2(Y)}\right)\right].
\end{equation}

It is interesting to see how we recover the forward limit $\vq\to 0$ recalled in 
Section \ref{sec:bindep}. This can be done 
remarking that formula \eqref{eq:kkq_case1}, obtained for BFKL dynamics in the 
case 1 ($k\gg k_0\gg q$), introduces and additional 
factor $(q/k_0)^{-2\gamma}$. By recombination of the $q/k$ and $q/k_0$ factors, 
a different factorisation appear where $k_0$ 
substitute to $q$ as the reference scale, as clearly seen in 
\eqref{eq:kkq_case10}. Inserting \eqref{eq:kkq_case10} in 
\eqref{eq:bkfkq}, by straightforward algebra, it is easy to see that the result 
can be cast under the form
\[
f(\vk,\vq\to 0) = \frac{1}{k^2} 
\int\frac{d\gamma}{2i\pi}\,\phi^\gamma(Q_T)\,e^{\bar\alpha\chi(\gamma)Y-\gamma 
L_0},
\]
where
\[
L_0=\log\left(\frac{k^2}{Q_T^2}\right),
\]
and $Q_T$ is a scale typical for the target, defining the initial condition for 
the traveling-wave solution. This corresponds to the 
solutions of equation \eqref{eq:bkb}.

This discussion can be translated to the amplitude in the mixed space 
representation of section \ref{sec:coord} defining similarly
\[
f(\vrh,\vq) = \int \frac{d^2\rho_0}{(2\pi)^2}\,\Phi_T(\vrp, 
\vq)\int\frac{d\gamma}{2i\pi}e^{\bar\alpha\chi(\gamma)Y}f_q^\gamma(\vrh, 
\vrp),
\]
where $f_q^\gamma(\vrh, \vrp)$, given by \eqref{eq:fqgamma}, is the solution of 
the BFKL equation in the mixed representation.
As for the case of the momentum space, the factorisation of the target and 
projectile dependences in the BFKL amplitude is the 
important property, leading to traveling waves when we include the nonlinear 
effects in the limit $\rho \ll 1/q$.

The situation is not as simple in coordinate space. Indeed, it is hard to 
introduce an amplitude depending only on $\vrh$ and $\cb$ 
from equation \eqref{eq:amplrrb}. This is closely related to the fact that 
$f^\gamma(\vrh, \vrp,\cb)$, depending only on the 
anharmonic ratio, cannot be factorised. In the limit where the BK equation is 
valid ($\rho_0\!\gg\!b,\rho$), its linear part gives 
solutions of the form (see \eqref{eq:rrbb}):
\[
f(\vrh,\cb)=\int\frac{d\gamma}{2i\pi}
\frac{2c_\gamma }{\gamma^2(1-\gamma)^2}\,e^{\bar\alpha\chi(\gamma)Y}\,
\left(\frac{4\rho}{\rho_0}\right)^{2\gamma},
\]
where $\rho_0$ is the typical size of the target. As explained in Section 
\ref{sec:bindep}, this leads to traveling-wave solutions 
for the BK equation ${\cal N}(\rho,Y)\!=\!{\cal N}(\rho\ Q_s(Y))$, where all 
trace of the impact parameter has disappeared. Hence, 
it gives no information upon the $b$ dependence. We could instead consider 
equation \eqref{eq:rrb_mean} when the scaling variable 
$\rho\rho_0/|\rho_0^2-4b^2|$ is large, {\em i.e.} $\rho_0\sim 2b$. However, it 
is not clear whether the BK equation makes sense 
physically in that limit where the dipole hits the target in its peripheral 
region ($\rho_0\sim 2b$). As explained above, these 
difficulties arise from the fact that $\rho$ and $\rho_0$ are mixed 
non-trivially in the anharmonic ratio \eqref{eq:ratio}. Hence, 
an interpretation in term of the BK equation is problematic. By contrast, the 
situation in mixed and momentum spaces does not suffer 
this inconvenient, due to the nice factorisation property in $\rho$ and 
$\rho_0$, or in $k$ and $k_0$. 

\section{Conclusions and perspectives}\label{sec:ccl}

Let us first emphasise the main results of this paper. We show that 
traveling-wave solutions of the forward BK equation can be 
extended to the full equation including nonzero transfer $q$, provided that 
$k\gg q$, where $k$ represents the scale of the 
projectile. The saturation scale \eqref{eq:newqs} has the same energy dependence 
than in the forward case but is now proportional to 
$q$. When this scale is large w.r.t. the scale, {\em e.g.} $Q_T$, characterising 
the target, the saturation scale thus becomes 
independent of the details of the target. But, when the momentum transfer $q$ is 
small w.r.t. the target scale, $Q_T$ substitutes to 
$q$ in the saturation scale as expected from the forward case analysis. 
Our results show that non-forward BKFL evolution with momemtum transfer $q$ in a 
presence of saturation at the scale $\Omega_s(Y)$ is equivalent to forward BKFL 
evolution in the presence of saturation at the scale $q\Omega_s(Y)/Q_T.$
Note that the similarity of a momentum transfer with an absorptive boundary in 
BFKL evolution has been noticed in Ref.\cite{mueltri}.

As a consequence of the existence of traveling-wave solutions, we derive the 
asymptotic form of the solution, {\em i.e.} the 
wavefront \eqref{eq:newfront}, at small $q/k$. It appears as an expression of 
the same form as in the forward case with two 
modifications: the saturation scale is now the proportional to $q$ and the 
pre-factor, related to the initial conditions, is now 
$q$-dependent.

From these considerations, we can predict that the BK equation implies the 
extension of geometric scaling at nonzero momentum 
transfer. Indeed, using formula \eqref{eq:amplkkq} and noting that, at large 
$k$, the impact factor of the projectile is expected to 
scale with the ratio $k/Q$ where $Q$ is a typical hard scale of the projectile. 
We see that the amplitude satisfies the geometric 
scaling under the form
\begin{equation}
{\cal{A}}(s,q^2,Q^2) \propto is\,\phi^{\gamma_c}(q) 
f\left(\frac{Q^2}{q^2\Omega^2_s(Y)}\right).
\end{equation}

All these results apply also in the mixed-space representation depending on the 
projectile size $\rho$ and the momentum transfer 
$q$. However, we find that these properties seem not to be easily expressed in 
terms of the impact parameter.

Technically, the method used consists in three main steps. Firstly, we analyse 
the form of the BFKL solutions in full phase space 
and in different pertinent limits. Secondly, we use conformal-invariant 
properties of the BFKL dynamics to look for factorised 
solutions which then lead to solutions of the linear part of the BK equation. 
Thirdly, we infer the traveling-wave solutions, 
induced by nonlinear effects, in the kinematical regions and for initial 
conditions where the universality properties of the BK 
equation apply.

Some comments are in order. Our formula \eqref{eq:newfront} suggests a solution 
to the puzzling problem of the compatibility of the 
BK equation with the confinement scale 
\cite{Kovner:2001bh,Ferreiro:2002kv,Bondarenko:2003ym,Golec-Biernat:2003ym,
Gotsman:2004ra,Ikeda:2004zp}. The key point is to address 
the problem in momentum space where the non-perturbative dependence is 
factorised as clearly seen from the factor 
$\phi^{\gamma_c}(q)$ in equation \eqref{eq:newfront}, which may be characterised 
by a confinement scale. Fourier-transforming this 
result back to impact parameter space will break this factorisation property. 

The properties of the BK solutions at nonzero momentum transfer suggest to 
analyse the high-energy behaviour of suitable experimental 
processes. The electroproduction of $\rho$-mesons at nonzero transfer has been 
already analysed in impact parameter space 
\cite{Munier:2001nr} and our solutions of the BK equation incite to perform the 
analysis in momentum space.

\begin{acknowledgments}
The authors would like to thank Rikard Enberg for encouraging discussions, 
Krzysztof Golec-Biernat for reading the manuscript and Henri Navelet for 
correcting a crucial sign error. G.S. is funded by the National Funds for 
Scientific Research (Belgium).
\end{acknowledgments}

\begin{appendix}

\section{Calculation of $f^\nu(\vk,\vkp,\vq)$}\label{ap:fourier}

In this appendix, we compute the Fourier transform of
\begin{equation}\label{eq:enuqdef}
E_q^\nu(\rho) = 2^{3\mu} \abs{q}^{-2\mu} \Gamma^2(1+\mu) 
\left[J_\mu\left(\frac{\bar q \rho}{4}\right)J_\mu\left(\frac{q 
\bar\rho}{4}\right) - J_{-\mu}\left(\frac{\bar q 
\rho}{4}\right)J_{-\mu}\left(\frac{q \bar\rho}{4}\right) \right],
\end{equation}
with $\mu = -i\nu$. It is of course sufficient to compute the following integral
\[
I = \int d\rho\,d\phi\,\rho^{2+\alpha} e^{i\rho 
v\cos(\phi)}J_\mu\left(\frac{\abs{q} 
\rho}{4}e^{i(\phi-\psi)}\right)J_\mu\left(\frac{\abs{q} 
\rho}{4}e^{i(\psi-\phi)}\right),
\]
where $v = \abs{k-q/2}$ and $\phi$ (resp. $\psi$) is the angle between $k-q/2$ 
and $\rho$ (resp. $q$). This integral has been 
regularised at infinity by introducing a factor $\rho^\alpha$ with 
$-2<\alpha<-3/2$. If we expand the Bessel functions in series and 
use
\[
\int_0^{2\pi} d\phi\,e^{i[\rho v \cos(\phi)-m\phi]} = 2\pi e^{i\pi m/2} J_m(\rho 
v),
\]
we find
\[
I = (2\pi)\sum_{j,k=0}^\infty 
\frac{\left(\frac{\abs{q}e^{i\psi}}{8}\right)^{\mu+2j}\left(
\frac{\abs{q}e^{-i\psi}}{8}\right)^{\mu+2k}}{j!\, k!\, \Gamma(1+\mu+j) 
\Gamma(1+\mu+k)} \int_0^\infty d\rho\,\rho^{2+\alpha+2\mu+2j+2k}J_{2(j-k)}
(\rho v).
\]
The integration over $\rho$ can be performed using
\[
\int_0^\infty dt\,t^{\beta-1}J_m(zt) = 
\frac{1}{2\pi}\left(\frac{2}{z}\right)^\beta\Gamma\left(\frac{\beta+m}{2}\right)
\Gamma\left(\frac{\beta-m}{2}\right)\sin\left[\left(
\frac{\beta-m}{2}\right)\pi\right].
\]
This leads to a factorisation of the integral in conformal blocs:
\[
I = \sin\left[\left(\frac{3+\alpha+2\mu}{2}\right)\pi\right] f(k,q)f(\bar k, 
\bar q),
\]
where, using the doubling formula,
\begin{eqnarray*}
f(k,q) & = & \sum_{j=0}^\infty 
\frac{1}{j!}\frac{\Gamma\left(\frac{3}{2}+\frac{\alpha}{2}+\mu+2j\right)}{\Gamma
(1+\mu+j)}\left(\frac{\abs{q} 
e^{i\psi}}{8}\right)^{\mu+2j}\left(\frac{2}{v}\right)^{\frac{3}{2}+\frac{\alpha}
{2}+\mu+2j} \\
 & = & \left(\frac{q}{4k-2q}\right)^\mu 
\left(\frac{2}{\abs{k-q/2}}\right)^{\frac{3}{2}+\frac{\alpha}{2}}
\frac{\Gamma\left(\frac{3}{2}+\frac{\alpha}{2}+\mu\right)}{\Gamma(1+\mu)}
\hyper\left(\frac{3}{4}+\frac{\alpha}{4}+\frac{\mu}{2},
\frac{5}{4}+\frac{\alpha}{4}+\frac{\mu}{2};1+\mu;\left(\frac{q}{2k-q}\right)^2
\right).
\end{eqnarray*}
The hypergeometric function can be simplified (see {\em e.g.} Ref.\cite{math}), 
and, taking $\alpha$ to zero, we finally obtain
\begin{equation}
I = -\cos(\mu\pi)\frac{\Gamma^2\left(\frac{3}{2}+\mu\right)}{\Gamma^2(1+\mu)}
\left(\frac{2}{\abs{k}}\right)^3 
\abs{\frac{q}{4k}}^{2\mu}\hyper\left(\frac{3}{2}+\mu,\frac{1}{2}+\mu;1+2\mu;
\frac{q}{k}\right)\hyper\left(\frac{3}{2}+\mu,\frac{1}{2}+\mu;1+2\mu;
\frac{\bar{q}}{\bar{k}}\right).
\end{equation}
The final result follows from adding to $I$ the corresponding expression with 
$\mu\to-\mu$ and using \eqref{eq:enuqdef}.

\section{Fourier transform of $E_q^\nu(\rho)$}\label{ap:sl2c}

In this appendix, we show how the Fourier transform of $E_q^\nu(\vrh)$ can be 
calculated using  conformal invariance. That will 
exhibit the link between $f^\nu(\vk,\vkp,\vq)$ and matrix elements of 
$SL(2,\mathbb{C})$ \cite{Janik:1999fk}. One has to deal with
\[
\int d^2\rho\, e^{i(\vk-\vq/2).\vrh} \frac{\abs{\vrh}}{4} E_q^\nu(\vrh)=
\frac{c_\gamma}{\pi^2}\int d^2\rho_1 d^2\rho_2\ 
e^{i\vq.\vrh_2}e^{i\vk.(\vrh_1-\vrh_2)}
|\vrh_1-\vrh_2|^{2\gamma}|\vrh_1|^{-2\gamma}|\vrh_2|^{-2\gamma},
\]
where $c_\gamma$ is given by \eqref{eq:cgamma}.

To compute this integral, one can switch to complex coordinates and perform the 
following changes of variables: first 
$\rho_2\!\rightarrow\!u\!=\!\rho_2/\rho_1,$ then 
$\rho_1\!\rightarrow\!w\!=\!\rho_1(\bar k\!+\!u(\bar q\!-\!\bar k))$, and 
finally 
$u\!\rightarrow\!v\!=\!u/(u\!-\!1).$ One obtains
\[
\int d^2\rho\, e^{i(\vk-\vq/2).\vrh} \frac{\abs{\vrh}}{4} E_q^\nu(\vrh)=
\frac{c_\gamma|\vk|^{2\gamma-4}}{\pi^2}\int d^2w\ e^{i Re(w)}|w|^{2-2\gamma}
\int d^2v |v|^{-2\gamma} |1\!-\!v|^{-2\gamma}|1\!-\!v\bar q/\bar 
k|^{2\gamma-4}\,,
\]
Then, using
\[
\int d^2w\ e^{i Re(w)}|w|^{2-2\gamma}=
\pi 2^{4-2\gamma}\frac{\Gamma(2-\gamma)}{\Gamma(\gamma-1)}\ .
\]
and the following result~\cite{gernav}
\begin{eqnarray*}
\int d^2v\ |v|^{2a_1-2} |1-v|^{2b_1-2a_1-2}|1-vx|^{-2a_0}=\pi
\left[\frac{\Gamma(a_1)\Gamma(b_1\!-\!a_1)\Gamma(1\!-\!b_1)}
{\Gamma(1\!-\!a_1)\Gamma(1\!-\!b_1\!+\!a_1)\Gamma(b_1)}  
{}_2 F_1(a_0,a_1;b_1;x){}_2 F_1(a_0,a_1;b_1;\bar x)+\right.\\
\left.+\frac{\Gamma(b_1\!-\!1)\Gamma(1\!-\!a_0)\Gamma(1\!-\!b_1\!+\!a_0)}
{\Gamma(2\!-\!b_1)\Gamma(a_0)\Gamma(b_1\!-\!a_0)}
|x|^{2\!-\!2b_1}{}_2 F_1(a_0\!-\!b_1\!+\!1,a_1\!-\!b_1\!+\!1;2\!-\!b_1;x)
{}_2 F_1(a_0\!-\!b_1\!+\!1,a_1\!-\!b_1\!+\!1;2\!-\!b_1;\bar x)\right],
\end{eqnarray*}
one easily recovers % the result \eqref{eq:fkkpq}.
\begin{eqnarray*}
\lefteqn{\int d^2\rho\, e^{i(\vk-\vq/2).\vrh} \frac{\abs{\vrh}}{4} 
E_q^\nu(\vrh)=
\frac{1}{4}\abs{q}^{2i\nu}2^{3-6i\nu}\Gamma^2(1-i\nu)\cos(i\nu\pi)\frac{1}
{\abs{k}^3}}\\
&&\left\lbrack\frac{\Gamma^2\left(\frac{3}{2}+i\nu\right)}{\Gamma^2(1+i\nu)} 
\abs{\frac{q}{4k}}^{2i\nu}\hyper\left(\frac{3}{2}+i\nu,\frac{1}{2}+i\nu;1+2i\nu;
\frac{q}{k}\right)\hyper\left(\frac{3}{2}+i\nu,\frac
{1}{2}+i\nu;1+2i\nu;\frac{\bar{q}}{\bar{k}}\right)-(\nu\to-\nu)\right\rbrack.
\end{eqnarray*}
The same integral appears in the calculation of matrix elements of 
$SL(2,\mathbb{C})$ \cite{Janik:1999fk} and more generally in 
conformal field theory \cite{gernav}.

\section{Asymptotic expansion at large $q$}\label{ap:asymp}

We want to emphasise the main steps yielding the asymptotic behaviour 
\eqref{eq:2f1exp}. We shall start by the asymptotic expansion 
of the hypergeometric function at infinity ($z = q/k \gg 1$)
\begin{eqnarray*}
\hyper(\gamma,\gamma+1;2\gamma;z) & = & 
\frac{2^{2\gamma-1}}{\sqrt{\pi}}\frac{\Gamma\left(\frac{1}{2}+\gamma\right)}
{\Gamma(1+\gamma)}(-z)^{-\gamma}
\left\{
1-\gamma(1-\gamma)\frac{\log(-z)}{z}
\hyper\left(\gamma+1,2-\gamma;2;\frac{1}{z}\right)\right.\\
&&\left.-\sum_{k=0}^\infty 
\frac{(\gamma)_{k+1}(1-\gamma)_{k+1}}{k!\,(k+1)!}z^{-k-1}
\left[
\psi(k+1)+\psi(k+2)-\psi(\gamma+k+1)-\psi(\gamma-k-1)
\right]
\right\}.
\end{eqnarray*}
When combining the $z$ and $\bar{z}$ parts, most of the pre-factors cancel:
\[
\frac{\Gamma^2(1+\gamma)}{\Gamma^2\left(\frac{1}{2}+\gamma\right)} 
\abs{\frac{z}{4}}^{2\gamma-1}\hyper(\gamma,\gamma+1;2\gamma;z)\hyper(\gamma,
\gamma+1;2\gamma;\bar{z}) = 
\frac{1}{\pi\abs{z}}\left[S_\gamma(z)+A_\gamma(z)\right]\left[S_\gamma(\bar 
z)+A_\gamma(\bar z)\right],
\] 
where
\begin{eqnarray*}
S_\gamma(z) & = & 
1-\gamma(1-\gamma)\frac{\log(-z)+\psi(1)+\psi(2)}{z}+{\cal{O}}(1/z^2), \\
A_\gamma(z) & = & 
\frac{1}{z}\left\{\gamma(1-\gamma)\left[\psi(\gamma+1)+\psi(\gamma-1)\right]+
(\gamma+1)\gamma(\gamma-1)(\gamma-2)\frac{\psi(\gamma+2)
+\psi(\gamma-2)}{z}+{\cal{O}}(1/z^2)\right\}.
\end{eqnarray*}
In this splitting we have put in $S_\gamma$ the terms which are invariant under 
the replacement $\gamma\to 1-\gamma$ and the 
remaining ones in $A_\gamma$. Therefore, if we consider the full $z$-dependent 
term, we easily get
\begin{eqnarray*}
\lefteqn{\frac{\Gamma^2(1+\gamma)}{\Gamma^2\left(\frac{1}{2}+\gamma\right)} 
\abs{\frac{z}{4}}^{2\gamma-1}\hyper(\gamma,\gamma+1;2\gamma;z)\hyper(\gamma,
\gamma+1;2\gamma;\bar{z}) - (\gamma\to 1-\gamma)}\\
& = & \frac{1}{\pi\abs{z}} \left\{
S_\gamma(z)\left\lbrack A_\gamma(\bar z)-A_{1-\gamma}(\bar z)\right\rbrack
+ S_\gamma(\bar z)\left\lbrack A_\gamma(z)-A_{1-\gamma}(z)\right\rbrack
+ \abs{A_\gamma(z)}^2 - \abs{A_{1-\gamma}(z)}^2\right\}.
\end{eqnarray*}
Obviously, the leading term in that development is proportional to 
$z^{-1}+\bar{z}^{-1}$, {\em i.e.} to $\frac{k}{q}+\frac{\bar 
k}{\bar q}$, which vanishes if we average over the angle between the two 
vectors. With less restrictions, one can also remark that 
this term is antisymmetric under the replacement $k\to -k$.

We thus go to the next order which finally gives a leading contribution of the 
form
\[
\frac{1}{\pi}\gamma^2(1-\gamma)^2\left[\psi(2-\gamma)+\psi(-\gamma)-\psi(\gamma+
1)-\psi(\gamma-1)\right]\frac{\log\left(\abs{z}^2\right)}{\abs{z}^3} = 
2\gamma^2(1-\gamma)^2\frac{\cos(\gamma\pi)}{\sin(\gamma\pi)}\frac{\log\left(
\abs{z}^2\right)}{\abs{z}^3}.
\]

\end{appendix}


\begin{thebibliography}{99}

\bibitem{Stasto:2000er}
A.~M. Sta\'sto, K.~Golec-Biernat, and J.~Kwiecinski,
\newblock Phys. Rev. Lett. {\bf 86}, 596 (2001) [arXiv:hep-ph/0007192].
%%CITATION = HEP-PH 0007192;%%

\bibitem{endnote17}{For a review on saturation and references, see A.H.~Mueller, 
``Parton saturation: An overview'', arXiv:hep-ph/0111244.}

\bibitem{Nikolaev:1991ja}
N.~N. Nikolaev and B.~G. Zakharov,
\newblock Z. Phys. {\bf C49}, 607 (1991);
%%CITATION = ZEPYA,C49,607;%%
\newblock A.~H. Mueller,
\newblock Nucl. Phys. {\bf B415}, 373 (1994).
%%CITATION = NUPHA,B415,373;%%

\bibitem{Balitsky}
I.~I.~Balitsky,
%``Operator expansion for high-energy scattering,''
Nucl.\ Phys.\ {\bf B463}, 99 (1996)
[arXiv:hep-ph/9509348];
%%CITATION = HEP-PH 9509348;%%

\bibitem{Kovchegov}
Y.~V. Kovchegov,
\newblock Phys. Rev. {\bf D60}, 034008 (1999) [arXiv:hep-ph/9901281];
%%CITATION = HEP-PH 9901281;%%
Y.~V. Kovchegov,
\newblock  Phys. Rev. {\bf D61}, 074018 (2000), [arXiv:hep-ph/9905214].
%%CITATION = HEP-PH 9905214;%%

%\cite{Munier}
\bibitem{Munier1}
S.~Munier and R.~Peschanski,
%``Geometric scaling as traveling waves,''
Phys.\ Rev.\ Lett.\  {\bf 91}, 232001 (2003)
[arXiv:hep-ph/0309177]; 
%%CITATION = HEP-PH 0309177;%%
%``Traveling wave fronts and the transition to saturation,''
Phys.\ Rev.\ {\bf D69}, 034008 (2004)
[arXiv:hep-ph/0310357]; 
%%CITATION = HEP-PH 0309177;%%
%``UNIVERSALITY AND TREE STRUCTURE OF HIGH-ENERGY QCD,''
Phys.\ Rev.\ {\bf D70}, 077503 (2004)
[arXiv:hep-ph/0310357]. 

%\cite{Munier:2001nr}
\bibitem{Munier:2001nr}
S.~Munier, A.~M.~Stasto and A.~H.~Mueller,
% ``Impact parameter dependent S-matrix for dipole proton scattering from
%diffractive meson electroproduction,''
Nucl.\ Phys.\ {\bf B603}, 427 (2001)
[arXiv:hep-ph/0102291].
%%CITATION = HEP-PH 0102291;%%
S.~Munier and S.~Wallon,
%``Geometric scaling in exclusive processes,''
Eur.\ Phys.\ J.\ {\bf C30} (2003) 359
[arXiv:hep-ph/0303211].
%%CITATION = HEP-PH 0303211;%%

%\cite{Kowalski:2003hm}
\bibitem{Kowalski:2003hm}
H.~Kowalski and D.~Teaney,
%``An impact parameter dipole saturation model,''
Phys.\ Rev.\ {\bf D68}, 114005 (2003)
[arXiv:hep-ph/0304189].
%%CITATION = HEP-PH 0304189;%%

%\cite{Bondarenko:2003ym}
\bibitem{Bondarenko:2003ym}
S.~Bondarenko, M.~Kozlov and E.~Levin,
%``QCD saturation in the semi-classical approach,''
Nucl.\ Phys.\ {\bf A727}, 139 (2003)
[arXiv:hep-ph/0305150].
%%CITATION = HEP-PH 0305150;%%
             
%\cite{Golec-Biernat:2003ym}
\bibitem{Golec-Biernat:2003ym}
K.~Golec-Biernat and A.~M.~Stasto,
%``On solutions of the Balitsky-Kovchegov equation with impact parameter,''
Nucl.\ Phys.\ {\bf B668}, 345 (2003)
[arXiv:hep-ph/0306279].
%%CITATION = HEP-PH 0306279;%%

%\cite{Gotsman:2004ra}
\bibitem{Gotsman:2004ra}
E.~Gotsman, M.~Kozlov, E.~Levin, U.~Maor and E.~Naftali,
% ``Towards a new global QCD analysis: Solution to the non-linear equation at
%arbitrary impact parameter,''
Nucl.\ Phys.\ {\bf A742}, 55 (2004)
[arXiv:hep-ph/0401021].
%%CITATION = HEP-PH 0401021;%%

%\cite{Ikeda:2004zp}
\bibitem{Ikeda:2004zp}
T.~Ikeda and L.~McLerran,
%``Impact parameter dependence in the Balitsky-Kovchegov equation,''
arXiv:hep-ph/0410345.
%%CITATION = HEP-PH 0410345;%%
 
%\cite{Ferreiro:2002kv}
\bibitem{Ferreiro:2002kv}
E.~Ferreiro, E.~Iancu, K.~Itakura and L.~McLerran,
%``Froissart bound from gluon saturation,''
Nucl.\ Phys.\ {\bf 1710}, 373 (2002)
[arXiv:hep-ph/0206241].
%%CITATION = HEP-PH 0206241;%%

%\cite{Kovner:2001bh}
\bibitem{Kovner:2001bh}
A.~Kovner and U.~A.~Wiedemann,
%``Nonlinear QCD evolution: Saturation without unitarization,''
Phys.\ Rev.\ {\bf D66}, 051502 (2002)
[arXiv:hep-ph/0112140];
Phys.\ Rev.\ {\bf D66}, 034031 (2002)
[arXiv:hep-ph/0204277];
Phys.\ Lett.\ {\bf B551}, 311 (2003)
[arXiv:hep-ph/0207335].
%%CITATION = HEP-PH 0207335;%%

\bibitem{KPP}
R.~A. Fisher,
\newblock Ann. Eugenics {\bf 7}, 355 (1937);
\newblock A.~Kolmogorov, I.~Petrovsky, and N.~Piscounov,
\newblock Moscou Univ. Bull. Math. {\bf A1}, 1 (1937).

\bibitem{Bramson}
M.~Bramson,
\newblock Memoirs of the American Mathematical Society {\bf 285} (1983).

\bibitem{ebert}
U. Ebert, W. van Saarloos,  
\newblock Physica {\bf D146}, 1 (2000) [arXiv:cond-mat/0003181];
For a review, see W. van Saarloos,
\newblock Phys. Rep. {\bf 386}, 29 (2003).

\bibitem{brunet}
E.~Brunet and B.~Derrida,
\newblock Phys. Rev. {\bf E56}, 2597 (1997). See in particular the appendix for 
a derivation of the traveling wave solutions.

\bibitem{Lipatov86} 
L.~N.~Lipatov,
%``The Bare Pomeron In Quantum Chromodynamics,''
Sov.\ Phys.\ JETP {\bf 63}, 904 (1986)
[Zh.\ Eksp.\ Teor.\ Fiz.\  {\bf 90}, 1536 (1986)].
%%CITATION = SPHJA,63,904;%%

\bibitem{NP} H.~Navelet and R.~Peschanski,
%``Conformal invariance and the exact solution of BFKL equations,''
Nucl.\ Phys.\ {\bf B507}, 353 (1997)
[arXiv:hep-ph/9703238].
%%CITATION = HEP-PH 9703238;%%

%\cite{Navelet:1997tx}
\bibitem{Navelet:1997tx}
H.~Navelet and S.~Wallon,
%``Onium onium scattering at fixed impact parameter: Exact equivalence  between
%the color dipole model and the BFKL pomeron,''
Nucl.\ Phys.\ {\bf B522}, 237 (1998)
[arXiv:hep-ph/9705296].
%%CITATION = HEP-PH 9705296;%%

\bibitem{LipatovReview}
L.~N.~Lipatov,
Phys.\ Rept.\ {\bf 286}, 131 (1997)
[arXiv:hep-ph/9610276].
%%CITATION = HEP-PH 9610276;%%

\bibitem{GLR}  L.V. Gribov, E.M. Levin and M.G. Ryskin, {Phys. Rep.} {\bf 
100}, 1 (1983).

\bibitem{mueltri}
A.~H.~Mueller and D.~N.~Triantafyllopoulos,
%``The energy dependence of the saturation momentum,''
Nucl.\ Phys.\ B {\bf 640} (2002) 331
[arXiv:hep-ph/0205167].
%%CITATION = HEP-PH 0205167;%%

\bibitem{math} I.~S.~Gradshtein and I.~M.~Ryzhik, {\em Tables of integrals, 
series and products}, 1980, Academic Press, New-York.

%\cite{Janik:1999fk}
\bibitem{Janik:1999fk}
R.~A.~Janik and R.~Peschanski,
%``Conformal invariance and {QCD} pomeron vertices in the 1/N(c) limit,''
Nucl.\ Phys.\ {\bf B549}, 280 (1999)
[arXiv:hep-ph/9901426].
%%CITATION = HEP-PH 9901426;%%

\bibitem{gernav} 
Vl.S. Dotsenko and V.A. Fateev, {\it Nucl. Phys.} {\bf B240}, 312 (1984);
J. Geronimo and H. Navelet, {\it J. Math. Phys.} {\bf 44}, 2293 (2003) 
[arXiv:math-ph/0003019. 

\end{thebibliography}
\end{document}